\documentclass[twocolumn,showpacs,preprintnumbers,amsmath,amssymb]{revtex4}
\usepackage[latin9]{inputenc}
\usepackage{amsmath}
\usepackage{amssymb}

\makeatletter
\@ifundefined{textcolor}{}
{%
 \definecolor{BLACK}{gray}{0}
 \definecolor{WHITE}{gray}{1}
 \definecolor{RED}{rgb}{1,0,0}
 \definecolor{GREEN}{rgb}{0,1,0}
 \definecolor{BLUE}{rgb}{0,0,1}
 \definecolor{CYAN}{cmyk}{1,0,0,0}
 \definecolor{MAGENTA}{cmyk}{0,1,0,0}
 \definecolor{YELLOW}{cmyk}{0,0,1,0}
}

\usepackage{graphicx}
\usepackage{bm}
\usepackage{epstopdf}
\usepackage{slashed}

\makeatother

\begin{document}

\title{The point-charge self-energy in a nonminimal Lorentz violating Maxwell
Electrodynamics}

\author{L. H. C. Borges}
\email{luizhenriqueunifei@yahoo.com.br}

\affiliation{UNESP - Campus de Guaratinguetá - DFQ, Avenida Dr. Ariberto Pereira
da Cunha 333, CEP 12516-410, Guaratinguetá, SP, Brazil}

\author{F. A. Barone}
\email{fbarone@unifei.edu.br}

\affiliation{IFQ - Universidade Federal de Itajubá, Av. BPS 1303, Pinheirinho,
Caixa Postal 50, 37500-903, Itajubá, MG, Brazil}

\author{A. F. Ferrari}
\email{alysson.ferrari@ufabc.edu.br}

\affiliation{Universidade Federal do ABC - UFABC, Avenida dos Estados, 5001, Santo
André, SP, 09210-580, Brazil}

\affiliation{Indiana University Center for Spacetime Symmetries, Indiana University,
Bloomington, Indiana 47405-7105}
\begin{abstract}
In this letter we study the self-energy of a point-like charge for
the electromagnetic field in a non minimal Lorentz symmetry breaking
scenario in a $n+1$ dimensional space time. We consider two variations
of a model where the Lorentz violation is caused by a background
vector $d^{\nu}$ that appears in a higher derivative interaction.
We restrict our attention to the case where $d^{\mu}$ is a time-like
background vector, namely $d^{2}=d^{\mu}d_{\mu}>0$, and we verify
that the classical self-energy is finite for any odd spatial
dimension $n$ and diverges for even $n$. We also make some
comments regarding obstacles in the quantization of the proposed model.
 
\end{abstract}
\maketitle
Lorentz symmetry violations have been systematically studied in the
past years in different scenarios encompassing low energy, nuclear
and high energy physics, astrophysics and others, thus providing a
very extensive view of possible physical effects arising from an assumed
violation of Lorentz symmetry, which may indicate some new physics
at very small length scales such as the Planck length\,\cite{SeventhCPT}.
As a result, besides a deeper understanding of the theoretical possibilities
involving the spontaneous breaking of such an essential symmetry in
our understanding of particle physics, also an extensive set of very
high precision bounds on Lorentz violating (LV) parameters have been
obtained\,\cite{datatables}. Most of these works have been done
in the context of the Standard Model Extension (SME)\,\cite{SME}
that incorporates in the Standard Model the full set of gauge-invariant,
renormalizable LV interactions. Understanding the SME as an effective
field theory that derives from some more fundamental theory at very
high energy, it becomes natural to incorporate also non-minimal terms,
i.e., those which are not renormalizable. As a matter of fact, a systematical
study of the non-minimal LV operators that may be added to the SME
(still maintaining gauge invariance) have begun to gain momentum more
recently\,\cite{nonminimalLV}. As non-minimal operators are expected
to be suppressed by powers of the very high energy scale of the fundamental
theory triggering the Lorentz violation, it is a general expectation
that their effects will be subleading comparing to those already present
in the minimal SME. Also, from the technical viewpoint, the
presence of additional derivatives in the non-minimal LV terms are
linked to possible issues with unitarity, given the presence of ghost
modes. However, in certain situations, a non-minimal LV operator
might induce some effect in low-energy physics which cannot be replicated
by any minimal LV operator of the SME. Therefore, one might entertain
the hope of finding new interesting phenomena related to the non-minimal
LV.

In a recent work\,\cite{axionlv}, some of us have exposed a relation
between non-minimal LV and axion physics, since a particular setting
of LV couplings, generated by some Lorentz violating high energy dynamics,
could contribute to the standard (Lorentz invariant) axion-photon
coupling. The complete modification induced in the Maxwell electrodynamics
by the LV background considered in\,\cite{axionlv} was calculated
in\,\cite{maxwellHD}, and this result opens up the opportunity for
investigating several aspects of photon physics that may be affected
by the particular LV couplings considered in these works. As a first
step in this direction, we explored in\,\cite{Fontes2} the effects
of one of the non-minimal terms found in\,\cite{maxwellHD}, to wit,
\begin{equation}
d^{\lambda}d_{\alpha}\partial_{\mu}F_{\nu\lambda}\partial^{\nu}F^{\mu\alpha}\thinspace,\label{eq:nonminimalLV}
\end{equation}
in the classical electromagnetic interaction between sources, such
as point charges, dipoles, lines of current and Dirac strings. Here,
$d^{\mu}$ is a constant vector parametrizing the LV. Some new physical
effects due the LV were unveiled, such as a spontaneous torque on
an isolated electric dipole. An extensive study of dimension
six LV operators, of which Eq.\,\eqref{eq:nonminimalLV} can be considered
a particular case, was recently reported in\,\cite{CasanaSchreck},
where the question of causality in such models is throughly discussed.
In this letter we present another consequence of the presence of the
interaction \eqref{eq:nonminimalLV} in classical electrodynamics,
more specifically the regularization of the self-energy of a point
charge in certain number of spatial dimensions.

The self-energy of an electrical charge is a celebrated problem in
classical electrodynamics, representing one of the early divergence
problems that faced theoretical physics in the beginning of the twentieth
century, since the self-energy diverges linearly with the ultraviolet
cutoff. Dirac's quantum theory of the electron improved matters, reducing
the divergence to a logarithmic one, and the problem of calculating
the electron self-energy remained a central one during the key years
of the development of the modern approach of quantum field theory
and the renormalization program. Some historical perspective on the
early attempts to solve the self-energy problem can be found in\,\cite{schweber}.
Podolsky\,\cite{Podolsky} and later Lee-Wick\,\cite{LW} discussed
a generalization of electrodynamics including higher derivatives in
which no ultraviolet divergences appeared, in particular in the electron
self-energy. Recently, some of us studied in detail the finiteness
of the electron self-energy in the context of the Lee-Wick electrodynamics\,\cite{LWautoenergia},
showing that a finite result can be obtained for odd number of spatial
dimensions only. In this letter, we will show that a similar behavior
occurs for the electrodynamics modified by the non-minimal LV coupling
\eqref{eq:nonminimalLV}.

We consider first the simplest extension of the Maxwell theory incorporating
the LV term given in Eq.\,\eqref{eq:nonminimalLV}, which is defined
by the following Lagrangian density in $n+1$ dimensional space-time,
\begin{align}
{\cal L}_{(1)}= & -\frac{1}{4}F_{\mu\nu}F^{\mu\nu}-\frac{1}{2\gamma}\left(\partial_{\mu}A^{\mu}\right)^{2}+\frac{1}{2}d^{\lambda}d_{\alpha}\partial_{\mu}F_{\nu\lambda}\partial^{\nu}F^{\mu\alpha}\nonumber \\
 & +J^{\mu}A_{\mu}\ ,\label{eq:1}
\end{align}
where $A^{\mu}$ is the electromagnetic field coupled to an external
source $J^{\mu}$, with the associated field strength $F^{\mu\nu}=\partial^{\mu}A^{\nu}-\partial^{\nu}A^{\mu}$.
Here, $\gamma$ is a gauge fixing parameter, and $d^{\lambda}$ is
the background vector parametrizing the LV in our model.

The parameter $d^{\lambda}$ has inverse of mass dimension and is
taken to be constant and uniform in the reference frame where the
calculations are performed. Assumedly $d^{\mu}$ is of the order of
the inverse of some very large mass related to some fundamental theory
at the Planck scale from which the LV originates. As in\,\cite{Fontes2},
we restrict ourselves to the case of $d^{\mu}$ being a time-like
background vector, namely 
\begin{eqnarray}
d^{2}=d^{\mu}d_{\mu}=(d^{0})^{2}-{\bf {d}}^{2}>0\ ,\label{dmu}
\end{eqnarray}
where ${\bf {d}}=\left(d^{1},d^{2},\ldots,d^{n}\right)$. For other
classes of $d^{\mu}$ vectors we are not able to perform the necessary
integrals, so our results do not extend to those cases.

The propagator for our model, in the Feynman gauge $\gamma=1$, is
given by 
\begin{align}
D^{\mu\nu}(x,y) & =\int\frac{d^{n+1}p}{(2\pi)^{n+1}}\Biggl\{-\frac{\eta^{\mu\nu}}{p^{2}}+\frac{1}{[1-d^{2}p^{2}+(p\cdot d)^{2}]}\nonumber \\
 & \times\Biggl[-d^{\mu}d^{\nu}-\frac{(p\cdot d)^{2}}{p^{4}}p^{\mu}p^{\nu}\nonumber \\
 & +\frac{(p\cdot d)}{p^{2}}(p^{\mu}d^{\nu}+d^{\mu}p^{\nu})\Biggr]\Biggr\} e^{-ip\cdot(x-y)}\ .\label{eq:propagator}
\end{align}
Since the Lagrangian (\ref{eq:1}) is quadratic in the field variables
$A^{\mu}$, the vacuum energy due to the presence of the external
source is given by 
\begin{equation}
E=\frac{1}{2T}\int\int d^{n+1}x\ d^{n+1}yJ_{\mu}(x)D^{\mu\nu}(x,y)J_{\nu}(y)\ ,\label{energy1}
\end{equation}
where $T$ is the time variable, and the limit $T\to\infty$ is implicit\,\cite{BaroneHidalgo}.

We consider the external source $J$ corresponding to a point-like
stationary charge $q$ placed at position ${\bf {a}}=\left(a^{1},a^{2},\ldots,a^{n}\right)$,
\begin{equation}
J^{\mu}({\bf {x}})=q\eta^{\mu0}\delta^{n}\left({\bf {x}}-{\bf {a}}\right)\ ,\label{source}
\end{equation}
where $\delta$ is the Dirac delta function in $n$ spatial dimensions.
Substituting (\ref{source}) in (\ref{energy1}), using the explicit
form of the propagator in Eq. (\ref{eq:propagator}) and computing
the integrals in the following order: $d^{n}{\bf x}$, $d^{n}{\bf y}$,
$dx^{0}$, $dp^{0}$ and $dy^{0}$, we obtain 
\begin{align}
E_{(1)}= & \frac{q^{2}}{2}\left[\int\frac{d^{n}{\bf p}}{(2\pi)^{n}}\frac{1}{{\bf p}^{2}}\right.\nonumber \\
 & \left.-\frac{(d^{0})^{2}}{d^{2}}\int\frac{d^{n}{\bf p}}{(2\pi)^{n}}\frac{1}{\left({\bf p}^{2}+\frac{\left({\bf d}\cdot{\bf p}\right)^{2}}{d^{2}}\right)+\left(\frac{1}{d}\right)^{2}}\right]\ ,\label{Ener2EM}
\end{align}
where $d=\sqrt{d^{2}}$.

In order to calculate the second integral in Eq.\,(\ref{Ener2EM}),
we shall carry out a change in the integration variables in $n$ spatial
dimensions, in the same way as in\,\cite{Fontes}. First we split
the vector ${\bf {p}}=\left(p^{1},p^{2},\ldots,p^{n}\right)$ as follows
\begin{equation}
{\bf p}={\bf p}_{\perp}+{\bf p}_{\parallel}\thinspace,
\end{equation}
where the vectors ${\bf p}_{\parallel}$ and ${\bf p}_{\perp}$ are
respectively parallel and perpendicular to the vector ${\bf d}$,
i.e. 
\begin{equation}
{\bf p}_{\parallel}={\bf d}\Bigl(\frac{{\bf d}\cdot{\bf p}}{{\bf d}^{2}}\Bigr),\ \ {\bf p}_{\perp}={\bf p}-{\bf d}\Bigl(\frac{{\bf d}\cdot{\bf p}}{{\bf d}^{2}}\Bigr)\thinspace.\label{mudan1EM}
\end{equation}
We also define the vector ${\bf u}$ as follows, 
\begin{eqnarray}
{\bf u} & = & {\bf p}_{\perp}+{\bf p}_{\parallel}\sqrt{1+\frac{{\bf d}^{2}}{d^{2}}}\label{defq}\\
 & = & {\bf p}+{\bf d}\Bigl(\frac{{\bf d}\cdot{\bf p}}{{\bf d}^{2}}\Bigr)\left(\frac{\mid d^{0}\mid}{d}-1\right)\ .
\end{eqnarray}
With the previous definitions, we can write 
\begin{equation}
{\bf p}_{\parallel}=\frac{{\bf d({\bf d\cdot{\bf u)}}}}{{\bf d}^{2}}\frac{d}{\mid d^{0}\mid}\ ,\ {\bf p}_{\perp}={\bf u-\frac{{\bf d({\bf d\cdot{\bf u)}}}}{{\bf d}^{2}}}\thinspace,\label{mudan6EM}
\end{equation}
which implies in 
\begin{equation}
{\bf p}={\bf u}+\frac{({\bf {d}\cdot{\bf u}){\bf {d}}}}{{\bf d}^{2}}\left(\frac{d}{\mid d^{0}\mid}-1\right)\thinspace,\label{mudan22EM}
\end{equation}
and 
\begin{equation}
{\bf u}^{2}={\bf p}^{2}+\frac{({\bf d}\cdot{\bf p})^{2}}{d^{2}}\ .\label{zxc2}
\end{equation}
The Jacobian of the transformation from ${\bf p}$ to ${\bf u}$ can
be obtained from Eq.\,(\ref{mudan22EM}), resulting in 
\begin{equation}
\det\left[\frac{\partial{\bf {p}}}{\partial{\bf {u}}}\right]=\frac{1}{\sqrt{1+\frac{{\bf d}^{2}}{d^{2}}}}=\frac{d}{\mid d^{0}\mid}\ .\label{mudan5EM}
\end{equation}
With this change of variables, we obtain 
\begin{align}
E_{(1)}= & \frac{q^{2}}{2}\left[\int\frac{d^{n}{\bf p}}{(2\pi)^{n}}\frac{1}{{\bf p}^{2}}\right.\nonumber \\
 & \left.-\frac{\mid d^{0}\mid}{d}\int\frac{d^{n}{\bf u}}{(2\pi)^{n}}\frac{1}{{\bf {u}}^{2}+\left(\frac{1}{d}\right)^{2}}\right]\ .\label{Ener52EM}
\end{align}
Both integrals in (\ref{Ener52EM}) are performed along the same $n$-dimensional
space. To avoid misunderstandings, we rewrite them in the ${\bf k}$
variable, instead of ${\bf u}$ and ${\bf p}$, as follows 
\begin{align}
E_{(1)}= & \frac{q^{2}}{2}\int\frac{d^{n}{\bf k}}{(2\pi)^{n}}\Biggl[\frac{1}{{\bf k}^{2}}-\frac{\mid d^{0}\mid}{d}\frac{1}{{\bf {k}}^{2}+\left(\frac{1}{d}\right)^{2}}\Biggr]\label{energy3}\\
= & \frac{q^{2}}{2}\left[\left(1-\frac{\mid d^{0}\mid}{d}\right)\int\frac{d^{n}{\bf k}}{(2\pi)^{n}}\frac{1}{{\bf k}^{2}+\frac{1}{d^{2}}}\right.\nonumber \\
 & \left.+\frac{1}{d^{2}}\int\frac{d^{n}{\bf k}}{(2\pi)^{n}}\frac{1}{{\bf {k}}^{2}\left({\bf k}^{2}+\frac{1}{d^{2}}\right)}\right]\ .
\end{align}
Integrating in $n$-dimensional spherical coordinates, using that
the integral in the solid angle of ${\bf {k}}$ gives $2\pi^{n/2}/\Gamma\left(n/2\right)$,
we arrive at 
\begin{align}
E_{(1)}= & \frac{q^{2}}{\left(4\pi\right)^{n/2}\Gamma\left(n/2\right)}\left[\left(1-\frac{\mid d^{0}\mid}{d}\right)\int_{0}^{\infty}dk\frac{k^{n-1}}{k^{2}+\frac{1}{d^{2}}}\right.\nonumber \\
 & \left.+\frac{1}{d^{2}}\int_{0}^{\infty}dk\frac{k^{n-3}}{k^{2}+\frac{1}{d^{2}}}\right]\ ,
\end{align}
with $\Gamma$ standing for the Euler Gamma function. The remaining
integral can be performed by means of the formula 
\begin{equation}
\int_{0}^{\infty}dr\frac{r^{\beta}}{(r^{2}+C^{2})^{\alpha}}=\frac{\Gamma\Bigl(\frac{1+\beta}{2}\Bigr)\Gamma\Bigl(\alpha-\frac{(1+\beta)}{2}\Bigr)}{2(C^{2})^{\alpha-(1+\beta)/2}\Gamma(\alpha)}\thinspace,\label{formula}
\end{equation}
leading to 
\begin{align}
E_{(1)}= & \frac{q^{2}}{2^{n+1}\pi^{n/2}d^{n-2}}\Biggl[\left(1-\frac{\mid d^{0}\mid}{d}\right)\Gamma\left(1-\frac{n}{2}\right)\nonumber \\
 & +\frac{\Gamma\left(\frac{n}{2}-1\right)\Gamma\left(2-\frac{n}{2}\right)}{\Gamma\left(\frac{n}{2}\right)}\Biggr]\ .
\end{align}
This expression can be further simplified by using the basic properties
of the Gamma function leading to 
\begin{equation}
E_{(1)}=-\frac{q^{2}}{2^{n+1}\pi^{n/2}}\frac{\mid d^{0}\mid}{d^{n-1}}\Gamma\left(1-\frac{n}{2}\right)\ .\label{energy6}
\end{equation}

It is interesting to notice that the self energy of a point charge
in this theory is finite for odd $n$ and diverges for even $n$.
For instance, we have the following results for $n=1,3,5,7$, \begin{subequations}
\begin{eqnarray}
E_{(1)}\left(n=1\right) & = & -\frac{q^{2}}{4}\mid d^{0}\mid\ ,\\
E_{(1)}\left(n=3\right) & = & \frac{q^{2}}{8\pi}\frac{\mid d^{0}\mid}{d^{2}}\ ,\\
E_{(1)}\left(n=5\right) & = & -\frac{q^{2}}{48\pi^{2}}\frac{\mid d^{0}\mid}{d^{4}}\ ,\\
E_{(1)}\left(n=7\right) & = & \frac{q^{2}}{480\pi^{3}}\frac{\mid d^{0}\mid}{d^{6}}\ ,
\end{eqnarray}
\end{subequations}while for $n=2,4,6$ we have \begin{subequations}
\begin{eqnarray}
E_{(1)}\left(n=2\right) & = & \frac{q^{2}}{8\pi}\frac{\mid d^{0}\mid}{d^{2}}\Biggl[\frac{2}{n-2}+\gamma\Biggr]\ ,\\
E_{(1)}\left(n=4\right) & = & -\frac{q^{2}}{32\pi^{2}}\frac{\mid d^{0}\mid}{d^{3}}\Biggl[\frac{2}{n-4}+\gamma-1\Biggr]\ ,\\
E_{(1)}\left(n=6\right) & = & \frac{q^{2}}{128\pi^{3}}\frac{\mid d^{0}\mid}{d^{6}}\Biggl[\frac{1}{n-6}+\frac{\gamma}{2}-\frac{3}{4}\Biggr]\ ,
\end{eqnarray}
\end{subequations}$\gamma=0.5772156649$ being the Euler constant.

The result obtained so far can be generalized for a slightly
different model, in which the same LV vector $d^{\mu}$ appears also
in a minimal LV operator. More concretely, we consider the model
defined by 
\begin{align}
{\cal L}_{(2)}= & -\frac{1}{4}F_{\mu\nu}F^{\mu\nu}-\frac{1}{2\gamma}\left(\partial_{\mu}A^{\mu}\right)^{2}-\frac{1}{2}\mu^{2}d^{\mu}d_{\nu}F_{\mu\lambda}F^{\nu\lambda}\nonumber \\
 & +\frac{1}{2}d^{\lambda}d_{\alpha}\partial_{\mu}F_{\nu\lambda}\partial^{\nu}F^{\mu\alpha}+J^{\mu}A_{\mu}\thinspace,\label{modelo2}
\end{align}
where $\mu$ a mass scale that is introduced by dimensional
reasons. For consistency with experimental observations, $\mu$ can
be considered as arbitrary, but not too large in order to keep the
dimensionless combination $\mu^{2}d^{\mu}d_{\nu}$ small. As in the
previous case, we shall consider only the case where $d^{2}>0$. Clearly,
in the limit $\mu\to0$, the Lagrangian (\ref{modelo2}) reduces to
(\ref{eq:1}) at the classical level, so in this sense this
model can be seen as a generalization of the first one.

Again choosing the Feynman gauge $\gamma=1$, and following similar
steps employed previously to deal with the $d^{\mu}$ appearing in
the non minimal coupling, we can arrive at the following expression
for the self energy of a steady point-like charge, 
\begin{align}
E_{(2)}= & \frac{q^{2}}{2}\int\frac{d^{n}{\bf p}}{(2\pi)^{n}}{\tilde{D}}_{(2)}^{00}(p^{0}=0,{\bf p})\nonumber \\
= & \frac{q^{2}}{2}\frac{1-\mu^{2}{\bf d}^{2}}{1+\mu^{2}d^{2}}\int\frac{d^{n}{\bf p}}{(2\pi)^{n}}\frac{1}{{\bf p}^{2}-\mu^{2}({\bf d}\cdot{\bf p})^{2}}\nonumber \\
 & -\frac{q^{2}}{2}\frac{(d^{0})^{2}}{1+\mu^{2}d^{2}}\int\frac{d^{n}{\bf p}}{(2\pi)^{n}}\frac{1}{d^{2}{\bf p}^{2}+({\bf d}\cdot{\bf p})^{2}+(1+\mu^{2}d^{2})}\thinspace,\label{zxc3}
\end{align}
where ${\tilde{D}}_{(2)}^{\mu\nu}(p)$ is the Fourier transform of
the propagator corresponding to the Lagrangian (\ref{modelo2}). It
is instructive to compare this expression with Eq.\,\eqref{Ener52EM},
noticing that the effect of the quadratic piece involving the $d^{\mu}$
contributes with the $1\pm\mu^{2}d^{2}$ factors and the new term
$\mu^{2}({\bf d}\cdot{\bf p})^{2}$ in the propagator.

We deal with this new factors and arrive at the final result with
the help of an additional change of variables. For the first integral
in (\ref{zxc3}), we perform the change of integration variable given
by 
\begin{equation}
{\bf p}\to{\bf k}={\bf p}+{\bf d}\frac{({\bf d}\cdot{\bf p})}{{\bf d}^{2}}\left[\sqrt{1-\mu^{2}{\bf d}^{2}}-1\right]\thinspace,
\end{equation}
such that 
\begin{equation}
{\bf k}^{2}={\bf p}^{2}-\mu^{2}({\bf d}\cdot{\bf p})^{2}\thinspace,
\end{equation}
and also 
\begin{equation}
\left|\frac{\partial{\bf p}}{\partial{\bf k}}\right|=\frac{1}{\sqrt{1-\mu^{2}{\bf d}^{2}}}\thinspace.
\end{equation}
For the second integral in (\ref{zxc3}), we use (\ref{defq}). Collecting
terms, and performing some simple manipulations, we can show that
the self-energy (\ref{zxc3}) is given by 
\begin{align}
E_{(2)}= & \frac{q^{2}}{2}\frac{1}{1+\mu^{2}d^{2}}\times\nonumber \\
 & \times\left[\sqrt{1-\mu^{2}{\bf d}^{2}}\int\frac{d^{n}{\bf k}}{(2\pi)^{n}}\left(\frac{1}{{\bf k}^{2}}-\frac{1}{{\bf k}^{2}+\frac{1+\mu^{2}d^{2}}{d^{2}}}\right)\right.\nonumber \\
 & +\left.\left(\sqrt{1-\mu^{2}{\bf d}^{2}}-\frac{|d^{0}|}{d}\right)\int\frac{d^{n}{\bf k}}{(2\pi)^{n}}\frac{1}{{\bf k}^{2}+\frac{1+\mu^{2}d^{2}}{d^{2}}}\right]\thinspace.
\end{align}
Integrating out in the solid angle, using Eq.\,(\ref{formula}) and
performing some simple manipulations, we obtain the self energy 
\begin{equation}
E_{(2)}=-\frac{q^{2}}{2^{n+1}\pi^{n/2}}\frac{|d^{0}|}{d^{n-1}}(1+\mu^{2}d^{2})^{n/2-2}\Gamma\left(1-\frac{n}{2}\right)\thinspace.\label{AE2}
\end{equation}
Again, for even $n$ the self energy is divergent, while it is finite
for odd $n$. As expected, in the limit $\mu\to0$, the energy (\ref{AE2})
goes to the result presented in Eq. (\ref{energy6}).

A similar pattern was found in the Lee-Wick electrodynamics\,\cite{LWautoenergia},
in which a Lorentz invariant, higher derivative modification is introduced
in QED to tame its divergences. We see that the LV parameter $d^{\mu}$
also acts as a regulator for the self energy of a point-like charge,
which in our case turns out to be finite, yet dependent on the particular
reference frame used to describe the measurement experiment,
as it is customary in Lorentz violating theories. 

The Lagrangian in Eq.\,\eqref{eq:1} was also considered
to describe the electromagnetic interaction between different kinds
of classical sources in\,\cite{Fontes2}. We remark, however, that
if this model is considered at the quantum level, with the Maxwell
field coupling to other quantum fields, such that radiative corrections
can appear, the presence of the LV coupling in Eq.\,\eqref{eq:nonminimalLV}
alone does not guarantee finiteness at the quantum level. Actually,
the problem of radiative corrections in LV model is a highly non trivial
one, that have been discussed extensively in the last years. There
are models in which specific LV interactions generate finite, well
defined corrections at one loop\,\cite{aether}, however, in general,
these corrections can be divergent (thus requiring some renormalization
mechanism), and even ambiguous\,\cite{JackiwKostelecky}. In fact,
the basic LV coupling considered by us in this paper, given in Eq.\,\eqref{eq:nonminimalLV},
was generated as a radiative correction arising from a fermion loop
in a specific LV model: we refer the reader to Refs\,\cite{axionlv,maxwellHD}
for an extensive discussion of these quantum corrections.

Another well known issue concerning higher derivative theories
has to do with the presence of classical instabilities and/or ghost
states, related to the presence of additional poles in the propagator.
These problems have been extensively discussed, for example, in connection
with the Lee-Wick electrodynamics\,\cite{LW}, see for example\,\cite{anselmi}
and references therein. Even with the ongoing discussion on how to
treat these issues from the theoretical point of view, a Lee-Wick
extension of the Standard Model was proposed\,\cite{LWSM1,LWSM2}
and a study of several phenomenological aspects ensued. Regarding
non minimal LV models, questions of stability and unitarity are also
non trivial, and have been discussed in different contexts\,\cite{Casana1,Belich,Scarpelli,Reyes1,Reyes2}.
There is no known general prescription to discern which non minimal
LV models can still have a consistent quantum formulation, free of
instabilities and unitary: each specific model has to be studied individually.
On general grounds, however, these additional poles are expected to
appear at very high mass scales, so from the point of view of effective
field theories, suitable for phenomenological considerations in low
energy (relative to the scale where Lorentz violation is generated,
which assumedly is near the Planck scale), they might be ignored\,\cite{nonminimalLV}.

Concerning our specific model, the propagator in Eq.\,\eqref{eq:propagator}
presents, besides the usual pole at $p^{2}=0$, additional poles at
the zeros of the function $\Psi\left(p\right)=1-d^{2}p^{2}+\left(p\cdot d\right)^{2}$.
More explicitly, we are interested in the solutions in the complex
$p^{\mu}$ plane for the equation 
\begin{equation}
\Psi\left(p\right)={\bf d}^{2}\left(p_{0}\right)^{2}-2\left({\bf p}\cdot{\bf d}\right)d_{0}p_{0}+1+d^{2}{\bf p}^{2}+\left({\bf p}\cdot{\bf d}\right)^{2}=0\thinspace.
\end{equation}
Taking into account that $d^{2}=\left(d_{0}\right)^{2}-{\bf d}^{2}>0$,
one might be tempted to choose the preferred frame in which $d^{\mu}=\left(d,{\bf 0}\right)$
to simplify the calculations, but then one would find the condition
$1+d_{0}^{2}{\bf p}^{2}=0$, independent of $p_{0}$, which cannot
define a consistent dynamics. Disregarding this particular choice
as anomalous, we set $d_{0}=\eta\left|{\bf d}\right|$ with $\eta>1$,
in which case $d_{0}/\left|{\bf d}\right|=\eta$ and $d^{2}/{\bf d}^{2}=\eta^{2}-1=\varepsilon>0$.
Then, we can solve the condition $\Psi\left(p\right)$ for $p_{0}$
as a function of ${\bf p}$ and $d^{\mu}$, obtaining two solutions
\begin{equation}
p_{0}^{\pm}=\left({\bf p}\cdot\hat{{\bf d}}\right)\eta\pm\left[\varepsilon\left({\bf p}\cdot\hat{{\bf d}}\right)^{2}-\varepsilon{\bf p}^{2}-1/\left|{\bf d}\right|^{2}\right]^{1/2}\thinspace,
\end{equation}
where $\hat{{\bf d}}={\bf d}/\left|{\bf d}\right|$. These are two
additional poles in the propagator, appearing as a consequence of
the LV term in Eq.\,\eqref{eq:nonminimalLV}, and they represent
instabilities in the theory: assuming $\left|{\bf d}\right|\sim1/M$,
where $M$ is a very high mass scale related to the origin of the
Lorentz violation (a natural assumption is $M\sim M_{\text{Planck}}$),
and also that $\left|{\bf p}\right|\ll M$, we can write 
\begin{equation}
p_{0}^{\pm}\approx\left({\bf p}\cdot\hat{{\bf d}}\right)\eta\pm iM\thinspace.
\end{equation}
This pole, corresponding to an imaginary energy, clearly signalizes
an instability in the theory. 

One might verify that relaxing the condition $\left|{\bf p}\right|\ll M$
does not solve this issue; also, the same general picture arises in
the generalized model shown in Eq.\,\eqref{modelo2}. As a result,
the definition of a consistent quantum theory starting from the classical
model considered in this letter depends on whether these unphysical
poles can somehow be removed from the theory, as in Lee-Wick electrodynamics\,\cite{anselmi}
or Standard Model\,\cite{LWSM1,LWSM2}, or in the LV models considered
in\,\cite{Reyes2}, and this is a question that deserves further
investigation.

In summary, the emergence of additional poles for propagators
in LV scenarios in theories with higher order derivatives, which may
jeopardize unitarity and/or stability, is a common problem and can
be a way to determine restrictions that must be imposed in theories
of this kind in order to quantize them\,\cite{RUV}. The presence
of problematic poles can also be used to distinguish between theories
that are feasible to be quantized and theories that must be taken
just in the classical context, or considered as effective theories
valid up so some scale, smaller than the characteristic scale of these
additional poles\,\cite{nonminimalLV}. 

Our main objective was to exhibit another instance where higher derivative
terms, in this case arising from a Lorentz violation, can act as physical
regulators for the classical self-energy of a point-like charge. Differently
from the Lee-Wick electrodynamics considered in\,\cite{BaroneHidalgo},
it is expected that, in a Lorentz violating model, the value of the
self-energy can depend on the reference frame where it is measured.
A general result for a moving charge with respect to the background
field is a much more complicated problem that deserves to be further
investigated.

\textbf{\medskip{}
}

\textbf{Acknowledgments.} This work was supported by Conselho Nacional
de Desenvolvimento Científico e Tecnológico (CNPq) and Fundação de
Amparo a Pesquisa do Estado de São Paulo (FAPESP), via the following
grants: CNPq 304134/2017-1 and FAPESP 2017/13767-9 (AFF), CNPq 311514/2015-4
(FAB), FAPESP 2016/11137-5 (LHCB).

\end{document}